\documentclass[journal]{IEEEtran}

\usepackage{xcolor}
\usepackage{algorithmicx}
\usepackage{algpseudocode}
\usepackage{hyperref}
\usepackage{amssymb}
\usepackage{multicol}
\usepackage{subcaption}
\usepackage{graphicx}
\usepackage{lipsum}
\usepackage{soul}
\usepackage{cite}
\usepackage{tabularx}
\usepackage{array}
\usepackage{multirow}
\usepackage{fancyhdr}
\usepackage{graphicx}
\usepackage{amsmath}

\newcolumntype{L}{>{\raggedright\arraybackslash}X}
\algrenewcommand\algorithmicindent{0.8em}%

\algtext*{EndIf}

\begin{document}

\title{Noise-Aware Generative Microscopic \\ Traffic Simulation}

\author{Anonymous Authors}
\author{Vindula Jayawardana$^{*}$, Catherine Tang$^{*}$, Junyi Ji, Jonah Philion, Xue Bin Peng, Cathy Wu
\thanks{$^{*}$Vindula Jayawardana and Catherine Tang contributed equally.}
\thanks{Vindula Jayawardana, Catherine Tang, and Cathy Wu are with the Massachusetts Institute of Technology, Cambridge, MA, USA.
\tt\small{vindula@mit.edu, cattang@mit.edu, cathywu@mit.edu}}
\thanks{Junyi Ji is with Vanderbilt University, Nashville, TN, USA.
\tt\small{junyi.ji@vanderbilt.edu}}
\thanks{Jonah Philion and Xue Bin Peng are with the NVIDIA Corporation, Santa Clara, CA, USA.
\tt\small{jphilion@nvidia.com, japeng@nvidia.com}}
}



\maketitle

\begin{abstract}

Accurately modeling individual vehicle behavior in microscopic traffic simulation remains a key challenge in intelligent transportation systems, as it requires vehicles to realistically generate and respond to complex traffic phenomena such as phantom traffic jams.  While traditional human driver simulation models like the Intelligent Driver Model offer computational tractability, they do so by abstracting away the very complexity that defines human driving. On the other hand, recent advances in infrastructure-mounted camera-based roadway sensing have enabled the extraction of vehicle trajectory data, presenting an opportunity to shift toward generative, agent-based models that learn to reproduce driving behaviors directly from data. Yet, a major bottleneck remains: most existing datasets are either overly sanitized or lack standardization, failing to reflect the noisy, imperfect nature of real-world sensing. Unlike data from vehicle-mounted sensors—which can mitigate sensing artifacts like occlusion through overlapping fields of view and sensor fusion— infrastructure-based sensors surface a messier, more practical view of challenges that traffic engineers face every day. To this end, we present the I-24 MOTION Scenario Dataset (I24-MSD)—a standardized, curated dataset designed to preserve a realistic level of sensor imperfection, embracing these errors as part of the learning problem rather than an obstacle to overcome purely from preprocessing. Drawing from noise-aware learning strategies in computer vision, we further adapt existing generative models in the autonomous driving community for I24-MSD with noise-aware loss functions. Our results show that such models not only outperform traditional baselines in realism but also benefit from explicitly engaging with, rather than suppressing, data imperfection. We view I24-MSD as a stepping stone toward a new generation of microscopic traffic simulation that embraces the real-world challenges and is better aligned with practical needs. The dataset can be found at \href{https://ct135.github.io/i24-msd/}{https://ct135.github.io/i24-msd/}.

\end{abstract}


\section{Introduction}
Simulating traffic is not merely a practical exercise in transportation and infrastructure planning - it is a deep system modeling challenge, one that tests our ability to represent multi-agent behavior under real-world constraints. Microscopic traffic simulation, in particular, offers a compelling lens: by modeling individual vehicles as autonomous agents responding to local observations, they make it possible to examine how fine-grained vehicle interactions give rise to system-level traffic dynamics~\cite{barcelo2010fundamentals}. These models contrast sharply with their macroscopic counterparts, which smooth over individual agent-level behavior in favor of aggregate flow. In doing so, microscopic traffic simulation opens the door to richer questions about car-following, lane-changing, and can reveal the formation, propagation, and dissipation of stop-and-go traffic waves~\cite{helbing2002micro}.

\begin{figure}
    \centering
    \includegraphics[width=1\linewidth]{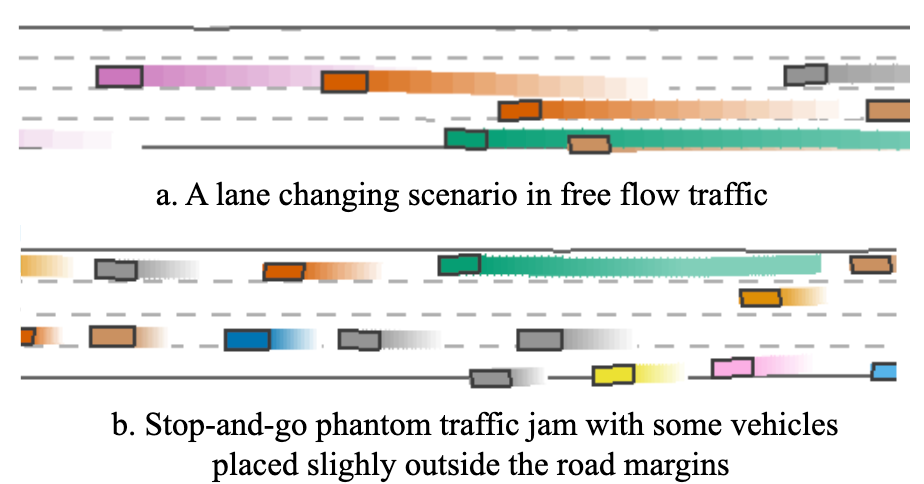}
    \caption{
        Example traffic scenarios from the I-24 MOTION Scenario Dataset of freeway driving in Interstate 24 in Nashville, Tennessee. Each bounding box represents a vehicle, with the shaded trail indicating its trajectory over the past second. Solid and dashed lines denote the road graph and corresponding lane boundaries, respectively. 
        \textbf{a.} A free-flow traffic scenario where vehicles change lanes without significantly affecting others. 
        \textbf{b.} A stop-and-go phantom traffic jam, where vehicles move slowly and intermittently. It also illustrates a data quality issue: some vehicles appear slightly misaligned with the lane markers due to road map annotation and/or multi-camera multi-object tracking issues. Illustrations are generated with MetaDrive~\cite{li2022metadrive}.
    }
    \label{fig:example}
\end{figure}

Yet, despite their conceptual richness, microscopic traffic simulation models have stagnated. Much of the field continues to rely on classical deterministic models like the Intelligent Driver Model (IDM)~\cite{treiber2013traffic}, which encodes human driving behavior as simplified differential equations focused on longitudinal control. These models, while analytically convenient, marginalize the full space of driving behavior—neglecting lateral maneuvers, multi-agent dependencies beyond the lead vehicle, and interactions shaped by road geometry or traffic control mechanisms. As a result, simulations built on them tend to miss the very dynamics that make real-world human driving a complex phenomenon, and hard to predict. They reflect, in essence, a kind of epistemic conservatism: holding on to what’s simple at the cost of what’s actually true. 

This classical landscape is shifting with the growing deployment of infrastructure-based sensing across major roadways that are producing vehicle trajectory data at scale. Datasets like those from I-24 MOTION in Tennessee~\cite{gloudemans202324}, DLR Highway Traffic Dataset in Germamy~\cite{schicktanz2025dlr}, and Zen Traffic Data in Japan~\cite{zentraffic} mark a turning point: they make it feasible to model traffic not as a system governed by deterministic rules, but as a learned distribution over agent behaviors conditioned on local context. In this view, simulation becomes a generative modeling problem—where each vehicle is an agent sampling actions autoregressively from a learned policy, grounded in past trajectory, neighboring agents, and the road map. This reframing is more than methodological; it changes the ontological commitment of simulation, shifting from deterministic modeling to probabilistic reproduction of behavior. The simulation no longer reflects what should happen according to a rule—it reflects what humans actually do, with all their inconsistencies, adaptations, and decision rationalities.

The challenge of data-driven traffic simulation from a generative modeling point of view is not new. The autonomous vehicle (AV) community has been grappling with it for a few years. Progress there has been accelerated by the release of high-fidelity, map-grounded, multi-agent trajectory datasets such as Waymo Open Motion~\cite{ettinger2021large}, Lyft Level 5~\cite{houston2021one}, nuScenes~\cite{caesar2020nuscenes}, and Argoverse~\cite{chang2019argoverse}, as well as by standardized benchmarks like the Waymo Sim Agents challenge~\cite{montali2023waymo}. We refer to these efforts as \textit{AV traffic simulation}. The parallels to microscopic traffic simulation are notable as both require learning generative models over agent behaviors. But while AV traffic simulation have built a thriving ecosystem around it, microscopic traffic simulation modeling remains largely out of sync.

Why the gap? We identify two core obstacles. First, infrastructure-based trajectory datasets are often released in raw, inconsistent formats, making it difficult to align them with modern generative simulation pipelines. Second, and more fundamentally, given the differences in the ways of collecting the data, the infrastructure-based data is noisy in ways that AV simulation data often is not—including errors originating from multi-camera, multi-object tracking, motion blur, suboptimal camera placement, and variations in lighting or weather~\cite{gloudemans202324,gloudemans2024so}. These artifacts make it nontrivial to learn robust generative models, especially those that depend on finely resolved agent-agent interactions. The challenge, then, is not just to scale modeling capacity, but to make it robust to data imperfections. 

Therefore, to move the field forward, we introduce the I-24 MOTION Scenario Dataset (I24-MSD), a structured, scenario-based dataset derived from the I-24 MOTION testbed~\cite{gloudemans202324}—the largest and most sensor-rich freeway monitoring system in the world~\cite{schicktanz2025dlr}. I24-MSD offers not only vehicle trajectories, but also aligned vectorized road maps, enabling spatially-aware human driver behavior modeling. Importantly, while the data has been processed using state-of-the-art techniques that are practical and accessible to transportation practitioners, it purposely retains some level of imperfections inherent to infrastructure-based sensing. This is by design: the dataset is intended not to abstract away noise, but to expose it—mirroring the real-world conditions under which models must ultimately operate. Its format is compatible with AV traffic simulation datasets, allowing for direct reuse of generative models and evaluation metrics in AV traffic simulation. This alignment is intentional: we see the AV traffic simulation community not as separate, but as a parallel research lineage that has simply advanced further down a shared path.

To explore this intersection concretely, we adapt SMART\cite{wu2024smart}, a state-of-the-art generative agent model originally developed for AV traffic simulation, to the microscopic traffic simulation with I24-MSD. Drawing inspiration from advances in vision and language modeling under noisy labels\cite{cambrin2024beyond}, we evaluate SMART’s performance using the standard cross-entropy loss and compare with three loss functions designed to mitigate label and context noise: (1) cross-entropy with label smoothing, (2) focal loss, and (3) symmetric cross-entropy. Across standard AV simulation metrics and classical microscopic traffic simulation baselines, we observe that incorporating loss robustness yields measurable gains—suggesting that imperfect infrastructure data, while challenging, need not be a barrier to learning effective generative traffic models.

Ultimately, we see this work as a vital link between AV traffic simulation and transportation research, sparking collaboration and driving progress on key challenges in microscopic traffic simulation.

\section{Microscopic Traffic Simulation as Conditional Generative Modeling}
\label{generative-traffic-simulations}

In this section, we formulate microscopic traffic simulation as a conditional generative modeling problem, drawing inspiration from formulations used in AV traffic simulation. We then highlight the key practical and objective differences between the two, emphasizing their distinct goals and technical constraints. To this end, we adapt the formulation presented in Montali et al.~\cite{montali2023waymo} and model the dynamics of the multi-vehicle microscopic traffic simulation using a Hidden Markov Model, defined as,

\begin{equation}
\mathcal{H} = \left( \mathcal{S}, \mathcal{O}, p(o_t \mid s_t), p(s_t \mid s_{t-1}) \right)
\end{equation}

\noindent where $\mathcal{S}$ is the set of latent world states and $\mathcal{O}$ is the space of observable state quantities. The emission distribution $p(o_t \mid s_t)$ specifies how observations are generated from the latent state at time $t$, while $p(s_t \mid s_{t-1})$ captures the Markovian dynamics of the underlying state transitions.

In the microscopic traffic simulation setting, we assume $N$ vehicle agents, and the observation at time $t$ is composed of their individual states, such as longitudinal and lateral positions and heading:

\begin{equation}
o_t = \left[o_t^{(1)}, o_t^{(2)}, \dots, o_t^{(N)}\right]
\end{equation}

\noindent where $o_t^{(i)}$ denotes the observed state of agent $i$ at time $t$.

The true observation dynamics are defined by marginalizing over the latent state sequence:

\begin{equation}
p_{\text{world}}(o_t \mid s_{t-1}) \triangleq \mathbb{E}_{p(s_t \mid s_{t-1})}[p(o_t \mid s_t)]
\end{equation}

The modeling objective of generative microscopic traffic simulation is to learn a generative world model $q_{\text{world}}(o_t \mid o^c_{<t})$ that approximates $p_{\text{world}}(o_t \mid s_{t-1})$ as closely as possible, using only observable state quantities. The conditioning context $o^c_{<t}$ consists of a static scene representation and a history of prior observations:

\begin{equation}
o^c_{<t} = \left[o_{\text{map}}, o_{\text{signs}}, o_{t-H-1}, \dots, o_{t-1}\right]
\end{equation}

\noindent where $H$ defines the observation history length, and $o_{\text{map}}$ and $o_{\text{signs}}$ denote static road map and traffic signs such as traffic signal and speed limits, respectively. The generative model $q_{\text{world}}$ must operate autoregressively for $T$ future time steps. 

In microscopic traffic simulation, observations are typically collected using infrastructure-mounted cameras. However, these recordings often suffer from noise and incompleteness due to factors such as occlusions, motion blur, and adverse visibility conditions. In contrast, AV traffic simulation rely on high-fidelity, vehicle-mounted sensors—such as lidar, radar, and high-resolution cameras. As a result, these observations are generally considered accurate proxies for the true latent state of the vehicles.

To formalize this difference, we extend the emission model to explicitly include an observation noise process. Let $o_t$ denote the true latent observable state of vehicles at time $t$, and $\tilde{o}_t$ the noisy observed version available to the model. The noisy observation is generated via a noise function:

\begin{equation}
\tilde{o}_t = \psi(o_t, \epsilon_t)
\end{equation}

\noindent where $\epsilon_t$ is a noise term, and $\psi$ is a function representing the corruption process (e.g., jitter, dropout, occlusion). This yields an updated $p_{\text{world}}(o_t \mid s_{t-1})$ model:

\begin{equation}
p_{\text{world}}(o_t \mid s_{t-1}) = \mathbb{E}_{p(s_t \mid s_{t-1})} \left[ \int p(\tilde{o}_t \mid s_t) \cdot p(o_t \mid \tilde{o}_t) \, d\tilde{o}_t \right]
\end{equation}

\noindent making explicit the role of observation noise.

This distinction gives rise to two different generative modeling paradigms. In AV traffic simulation, the generative model can be expressed as \( q_{\text{world}}^{\text{AV}}(o_t \mid o^c_{<t}) \approx q(o_t \mid \tilde{o}^c_{<t}) \), where the context \( \tilde{o}^c_{<t} \) can assumed to be noise-free or contain less noise and derived from richly annotated datasets. In contrast, microscopic traffic simulation require the model to operate under observation noise, and can be therefore formulated as \( q_{\text{world}}^{\text{micro-sim}}(o_t \mid o^c_{<t}) = q(o_t \mid \tilde{o}^c_{<t}= \psi({o}^c_{<t}, \epsilon_{<t})) \), where the context $\tilde{o}^c_{<t}$ is a noisy trajectory history after corruption process $\psi(\cdot)$.

When the generative model \( q_{\text{world}}^{\text{micro-sim}} \) is learned with parameters \( \theta \), the objective becomes Equation~\ref{obj} where $\mathcal{N}$ denotes the noise distribution, $\mathcal{D}$ denotes the dataset, and \( \mathcal{J} \) represents the loss function. 

\begin{equation}
\min_\theta \, \mathbb{E}_{\tilde{o}^c_{<t} \sim \mathcal{D}} \left[ \mathcal{J}\left(q^{\text{micro-sim}}_{\theta}(o_t \mid \tilde{o}^c_{<t} = \psi({o}^c_{<t}, \epsilon_{<t})), \tilde{o}_{t} \right) \right]
\label{obj}
\end{equation}

Then, the presence of observation noise in data can be treated with adjustments across multiple dimensions of the learning process. For example, robustness to noise can be introduced through: (1) the design of the loss function \( \mathcal{J} \), and (2) architectural choices in the model \( \theta \) that explicitly account for generalization. This may include noise-correcting supervised losses such as focal loss or symmetric cross-entropy, reinforcement learning objectives that penalize suboptimal behavior caused by noisy observations (e.g., imperfect driving decisions), or formulations that combine multiple such loss functions. Similarly, the model parameters \( \theta \) may reflect architectural choices that make the model robust to noise, such as incorporating uncertainty modeling, attention mechanisms focused on valid inputs, or dedicated denoising modules.

\subsection{The potential defining factors of $\psi(\cdot)$.}
\label{noise-sources}

The overall noise function $\psi(\cdot)$, encompasses a wide range of error sources that collectively degrade the fidelity of infrastructure-derived trajectory data. While modern systems rely on multi-camera setups and state-of-the-art computer vision algorithms for preprocessing, they remain fundamentally vulnerable to environmental and physical disturbances. For instance, thermal expansion of infrastructure poles under sunlight or tilting due to strong winds can induce subtle yet persistent shifts in camera orientation. These shifts degrade calibration accuracy over time and introduce spatial inconsistencies in trajectory projections—errors that are difficult to reverse without continuous ground truth access or dynamic recalibration systems, which are rarely available in practice.

Transient occlusions further contribute to data corruption. Dust, debris, motion blur, or nighttime glare can obscure the visual field, resulting in partial or total information loss. Unlike sensor noise, these occlusions often obliterate the signal entirely, rendering interpolation or imputation ineffective—particularly for subtle but behaviorally significant maneuvers such as lane changes, merges, or abrupt braking, which are critical in microscopic traffic simulation.

\begin{figure*}[ht!]
  \centering
  \begin{subcaptionbox}{The map of camera poles on Interstate 24\label{fig:sub1}}[0.44\textwidth]
    {\includegraphics[width=\linewidth]{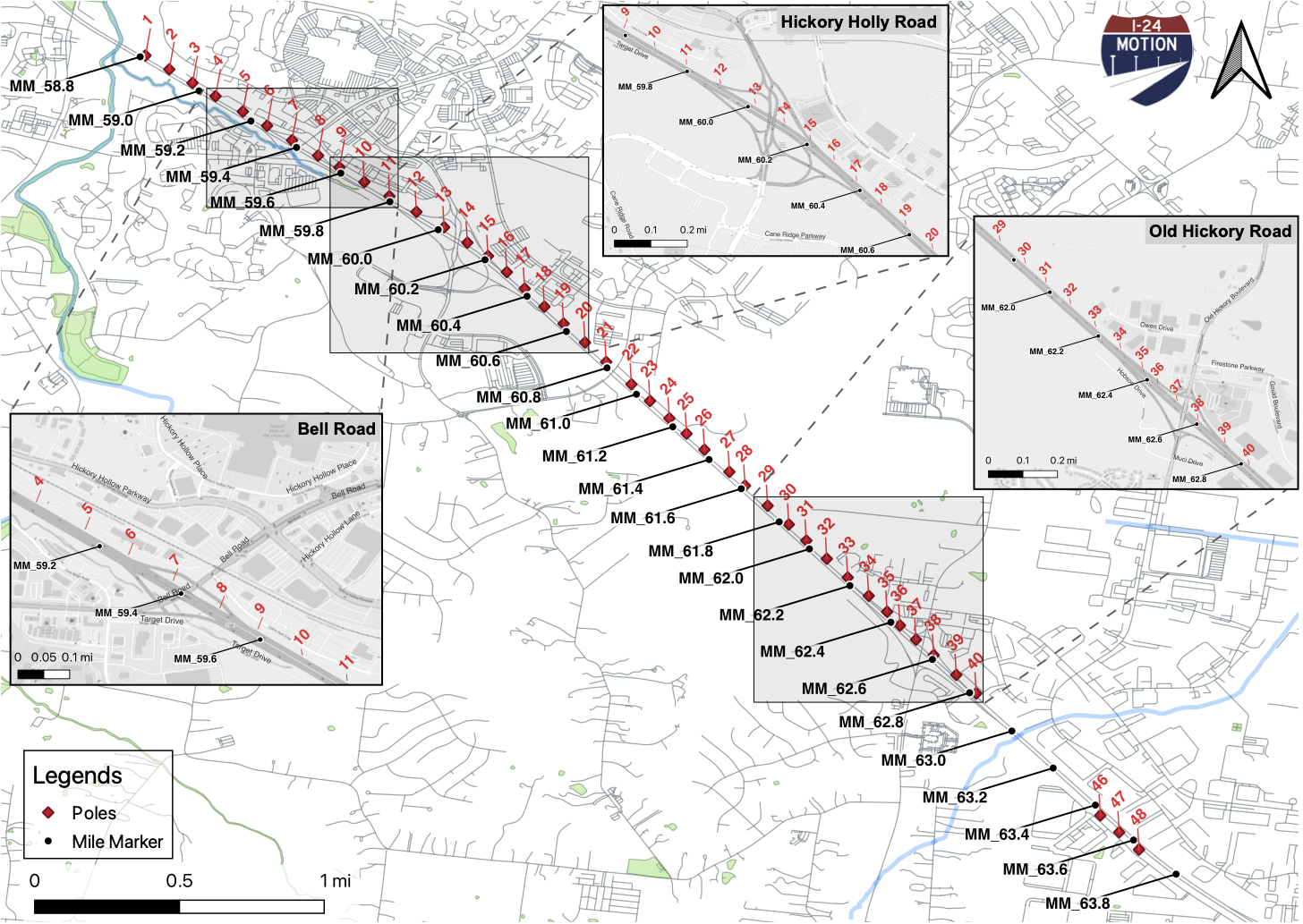}}
  \end{subcaptionbox}
  \hfill
  \begin{subcaptionbox}{Interstate 24 with 6 visible camera poles\label{fig:sub2}}[0.4\textwidth]
    {\includegraphics[width=\linewidth]{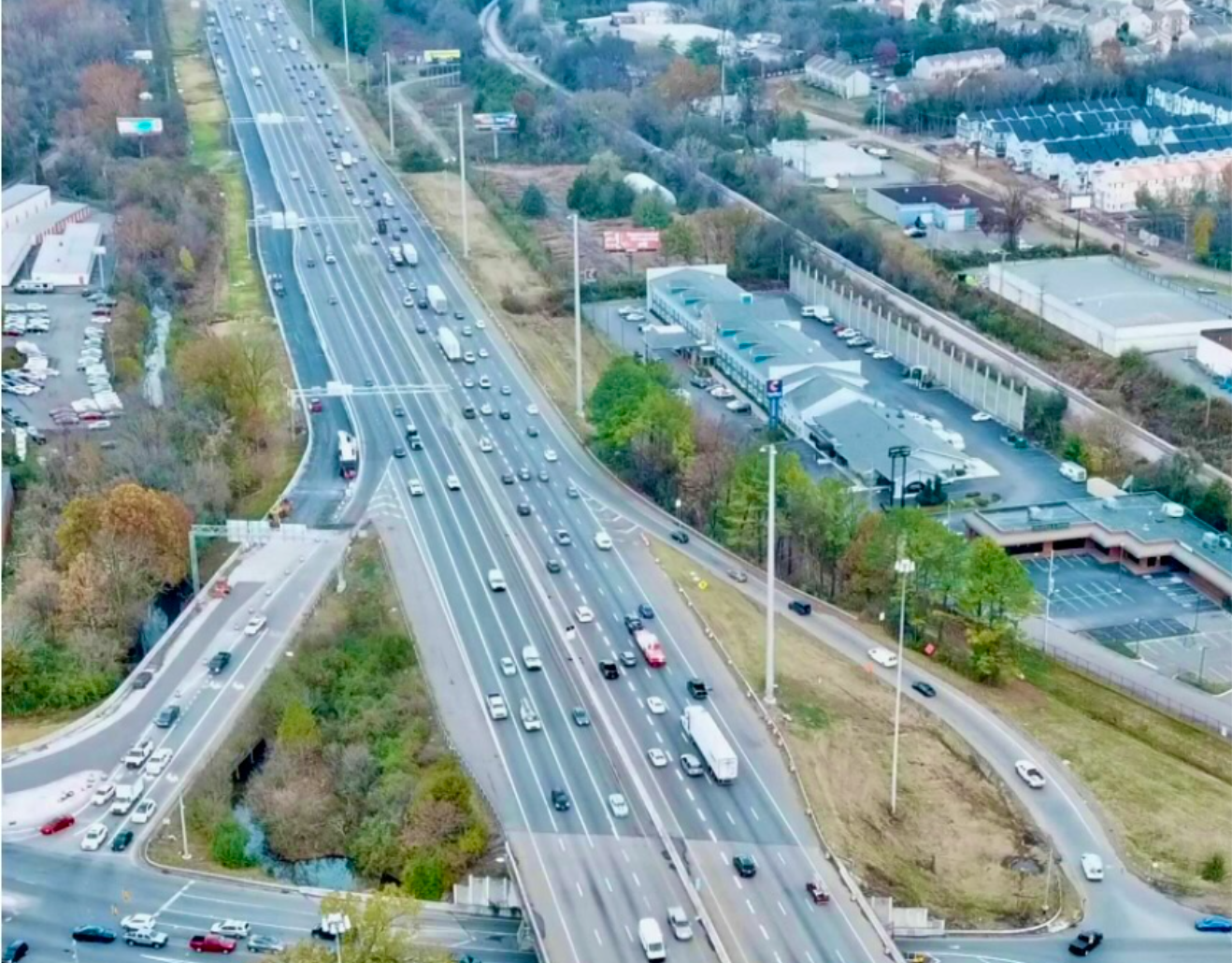}}
  \end{subcaptionbox}
  \hfill
  \begin{subcaptionbox}{Camera pole\label{fig:sub3}}[0.122\textwidth]
    {\includegraphics[width=\linewidth]{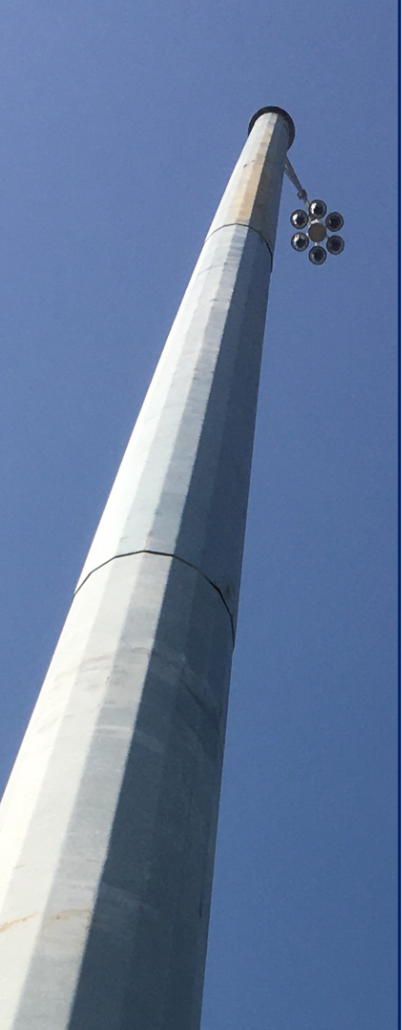}}
  \end{subcaptionbox}
  \caption{Illustrations of I-24 MOTION traffic testbed and infrastructure-based multi-camera system used to collect the data presented in I24-MSD. The figures are originally published and are taken from \cite{monache2025modeling, gloudemans202324}. }
  \label{fig:i24}
\end{figure*}

Hardware and system-level issues add another layer of complexity. Frame drops, corruption, or skipped captures—caused by firmware instability, bandwidth constraints, or network latency—create temporal discontinuities that fragment trajectory sequences. In multi-camera deployments, achieving consistent object tracking across overlapping fields of view requires precise timestamp alignment and robust identity matching, both of which are frequently undermined by desynchronized clocks, occlusions, or inconsistent detection performance. Common failure modes include ID switches, trajectory fragmentation, false negatives, and false positives—each compounding over time and varying unpredictably with traffic density, environmental conditions, and camera placement.

Additionally, road maps themselves often introduce alignment errors. Vectorized lane geometries may be imprecise or outdated, leading to spatial mismatches between observed trajectories and lane boundaries. Static assumptions about road signage ignore real-world dynamics such as temporary construction zones, lane closures, or accident-induced detours—all of which are reflected in the vehicle behaviors but not annotated in the datasets. 

The cumulative impact of these corruption sources manifests in a number of ways: noisy or jittered vehicle positions, inconsistent headings, missing or truncated trajectories, and mismatches between vehicle trajectories and the road maps. These inconsistencies could introduce a large number of outliers and edge cases into the dataset—driving behaviors that deviate significantly from the true underlying distribution of driving behavior. As a result, generative models tasked with learning realistic driver behavior could be easily misled. Instead of converging on a coherent representation of human driving, models often overfit to noise or fail to generalize across scenarios. The presence of spurious signals—such as lane changes into non-existent lanes, vehicles appearing to drive off-road, or implausible accelerations due to corrupted frame intervals—can distort loss landscapes during training and ultimately degrade predictive performance. This makes the learned models sensitive to artifacts rather than reflective of true behavioral dynamics, undermining both simulation realism and applicability to downstream applications. 

On the other hand, while the cross-domain transfer from AV traffic simulation methods to microscopic traffic simulation is an enticing prospect, we argue that a direct adaptation is unlikely to succeed due to the same reasons. The core obstacle lies in the stark contrast in data quality. AV traffic simulation datasets—such as Waymo Open Motion—are generated using cutting-edge, vehicle-mounted sensor suites that integrate lidar, radar, and high-resolution cameras, and are meticulously curated by large engineering teams to ensure precision and completeness. By contrast, infrastructure-based traffic datasets are typically collected using pole-mounted cameras under tighter budget and maintenance constraints. These fundamental differences in sensing modality and data curation result in significant disparities in data quality, resolution, coverage, and the richness of observable human driving behaviors~\cite{wang2024automatic, das2023comparison}—posing a major barrier to repurposing AV traffic simulation models without adaptation.

While it may be tempting to attribute these data quality challenges solely to shortcomings in processing pipelines, we argue that these imperfections are inherent to the broader challenge of microscopic traffic simulation. In other words, such limitations are not merely artifacts to be eliminated through improved preprocessing; rather, handling them is a fundamental requirement of the generative modeling problem itself. Treating them as peripheral issues—as has often been the case over the past decade—has significantly hindered progress in the field. Moreover, from a practical standpoint, expecting extensive data curation and high-end processing is often unrealistic, given the limited resources and tight budgets that constrain most transportation authorities. A more productive approach is to embrace these imperfections as constraints that generative models must learn to accommodate and operate within.

\section{Related work}

\subsection{Microscopic traffic simulation}

Microscopic traffic simulation have long been integral to traffic management and infrastructure planning. Classical tools such as SUMO~\cite{sumo}, VISSIM~\cite{vissim}, AIMSUN~\cite{aimsum}, and TransModeler~\cite{trans-modeler} have served as foundational platforms in transportation research and engineering. These simulations typically rely on car-following models like the Intelligent Driver Model (IDM)\cite{treiber2013traffic} and the Krauss model\cite{krauss1997metastable}, which are based on simplified differential equations and primarily consider interactions with the vehicle directly ahead. These traditional models remain widely used in both research and practice~\cite{wu2021flow, jayawardana2024mitigating}. Recent work has begun to explore data-driven approaches for capturing car-following behavior~\cite{wang2017capturing}, but such models often remain fail to account for more complex vehicle interactions. Consequently, the simplified nature of current microscopic traffic models has been shown to lead to inaccuracies in traffic flow analysis and predictions~\cite{ni2020limitations}.

\subsection{AV traffic simulation}

AV traffic simulation has become a critical component of the AV development pipeline. Its growth has been fueled by the availability of richly annotated datasets such as Waymo Open Motion~\cite{ettinger2021large}, Lyft Level 5~\cite{houston2021one}, nuScenes~\cite{caesar2020nuscenes}, and Argoverse~\cite{chang2019argoverse}, as well as simulation benchmarks like the Waymo Sim Agents Challenge~\cite{montali2023waymo}. Although the concept of AV simulation dates back to early efforts such as ALVINN~\cite{pomerleau1988alvinn}, the recent surge in AV research and the success of deep learning have significantly accelerated progress in this area.

A variety of generative approaches have emerged, including next-token prediction models~\cite{philion2023trajeglish, wu2024smart, zhang2024closed}, next-patch prediction models~\cite{zhou2024behaviorgpt}, and other transformer-based architectures~\cite{shi2022motion}. Additionally, variational autoencoders (VAEs)\cite{suo2021trafficsim}, generative adversarial networks (GANs)\cite{igl2022symphony}, and diffusion-based models~\cite{zhong2023guided} have been employed to improve the realism and diversity of simulated traffic behaviors.

\section{I24-MSD Dataset}

In this section, we introduce the I24 MOTION Scenario Dataset (I24-MSD)—a curated, standardized dataset designed to advance generative microscopic traffic simulation.

\subsection{Dataset creation}

I24-MSD is constructed from vehicle trajectory data collected at the I-24 MOTION testbed—the largest instrumented traffic monitoring system in the world~\cite{gloudemans202324}, located on Interstate 24 in Nashville, Tennessee. The dataset captures freeway driving behavior along a 4-mile westbound segment of I-24, recorded over 40 hours across 10 days. Figure~\ref{fig:i24} provides an overview of the testbed, including its location along the interstate, a photo of the infrastructure poles along the interstate, and the configuration of the pole-mounted cameras used to capture multi-vehicle trajectories.

We make the I24-MSD dataset compatible with popular AV traffic simulation datasets such as Waymo Open Motion by adopting the traffic scenario-based TFRecord format~\cite{tensorflow}. By default, each traffic scenario in I24-MSD contains up to 32 vehicle trajectories, each up to 9 seconds long, with a sample frequency of 10Hz. Apart from the vehicle trajectories, we also provide a vectorized road map that corresponds to that traffic scenario. The trajectories are provided as a sequence of x coordinate, y coordinate, z coordinate, and heading. The dataset is also released with the processing code to create custom datasets (defined by the maximum number of vehicles and the maximum length of a trajectory) as intended by the users, giving the flexibility for long-horizon trajectory prediction and many agent trajectory prediction. We set the current default maximum number of vehicles to 32 and the maximum length of a trajectory to 9 seconds to be compatible with existing generative models used in AV traffic simulation. 

For training and evaluation, I24-MSD offers predefined training, validation, and test splits. The training and validation sets contain naturally noisy data reflecting real-world conditions, while the test set is curated to reduce noise and serve as an approximate ground-truth reference.

\subsection{Processing traffic scenarios}

Since the I24-MSD dataset is built upon the I-24 MOTION data, we inherit the pre-processed vehicle trajectories from I-24 MOTION~\cite{gloudemans202324}. Gloudemans et al.~\cite{gloudemans202324} employed advanced post-processing techniques \cite{wang2024automatic,gloudemans2024so} from both computer vision and transportation research to extract these trajectories from infrastructure-mounted camera recordings. However, we observe a few limitations in the original dataset. This limitation stems from issues present in the original trajectories, including vehicle collisions, off-road vehicle positions, invalid movements, and fragmented trajectories. These challenges are inherent to infrastructure-based data collection and reflect the fundamental complexities of the task itself. To address these issues and enhance the dataset's suitability for microscopic traffic simulation, we apply a second stage of postprocessing.

As part of our second-stage postprocessing, we filter out trajectories whose positions fall entirely outside the road boundaries. However, we take care to preserve as many vehicles as possible to maintain a realistic traffic context. Completely off-road vehicles are removed, but those that merely graze the edges of the roadway—without fully departing from it—are retained to avoid creating unnatural or context-less driving scenarios. We further refine the dataset by filtering out trajectories exhibiting physically implausible behavior. This includes trajectories with excessively steep lateral movements—those that traverse multiple or all lanes within a short longitudinal distance—as well as trajectories that are unrealistically short. To eliminate crashing vehicle trajectories, or overlapping vehicles at the same timestep, we identify pairs of trajectories that are within one vehicle length of each other and remove the later-listed vehicle in such cases. Finally, for better map vectorization, we densify the original Interstate 24 map polylines by inserting 10 interpolated points between each pair of consecutive coordinates.

\textbf{Remark}: We note that the appropriate scope and nature of corrections and postprocessing should be performed remain in a grey area. Therefore, our objective here is to mimic the postprocessing steps that traffic engineers are most likely and able to perform, considering typical constraints in resources and expertise. The resulting dataset is thus intentionally crafted to retain a realistic level of noise and imperfections.

\subsection{Summary of the dataset}

\begin{table}[ht]
\centering
\caption{Comparison of datasets. The I24-MSD dataset is referred to as I24 for brevity. All entries in the table, except for I24, are taken directly from Ettinger et al.~\cite{ettinger2021large}. }
\label{dataset-comparison}
\resizebox{\columnwidth}{!}{%
\begin{tabular}{|l|c|c|c|c|c|c|}
\hline
 & Lyft & NuSc & Argo & Inter & Waymo & I24 \\
\hline
\# tracks & 53.4m & 4.3k & 11.7m & 40k & 7.64m & 3.29m \\
Avg len (s) & 1.8 & -- & 2.48 & 19.8 & 7.04 & 6.8 \\
Horizon (s) & 5 & 6 & 3 & 3 & 8 & 8 \\
\# segs & 170k & 1k & 324k & -- & 104k & 570k \\
Seg dur (s) & 25 & 20 & 5 & -- & 20 & 9 \\
Total hrs & 1118 & 5.5 & 320 & 16.5 & 574 & 40 \\
Roadways & 10km & -- & 290km & -- & 1750km & 6.5km \\
\hline
Rate (Hz) & 10 & 2 & 10 & 10 & 10 & 10 \\
Cities & 1 & 2 & 2 & $6^{*}$ & 6 & 1 \\
Obj types & 3 & $1^{\dagger}$ & $1^{\ddagger}$ & 1 & 3 & 1 \\
\hline
\end{tabular}
}
\end{table}

\begin{figure*}[t]
  \centering
  \begin{subcaptionbox}{Number of vehicles per scenario\label{fig:sub1}}[0.32\textwidth]
    {\includegraphics[width=\linewidth]{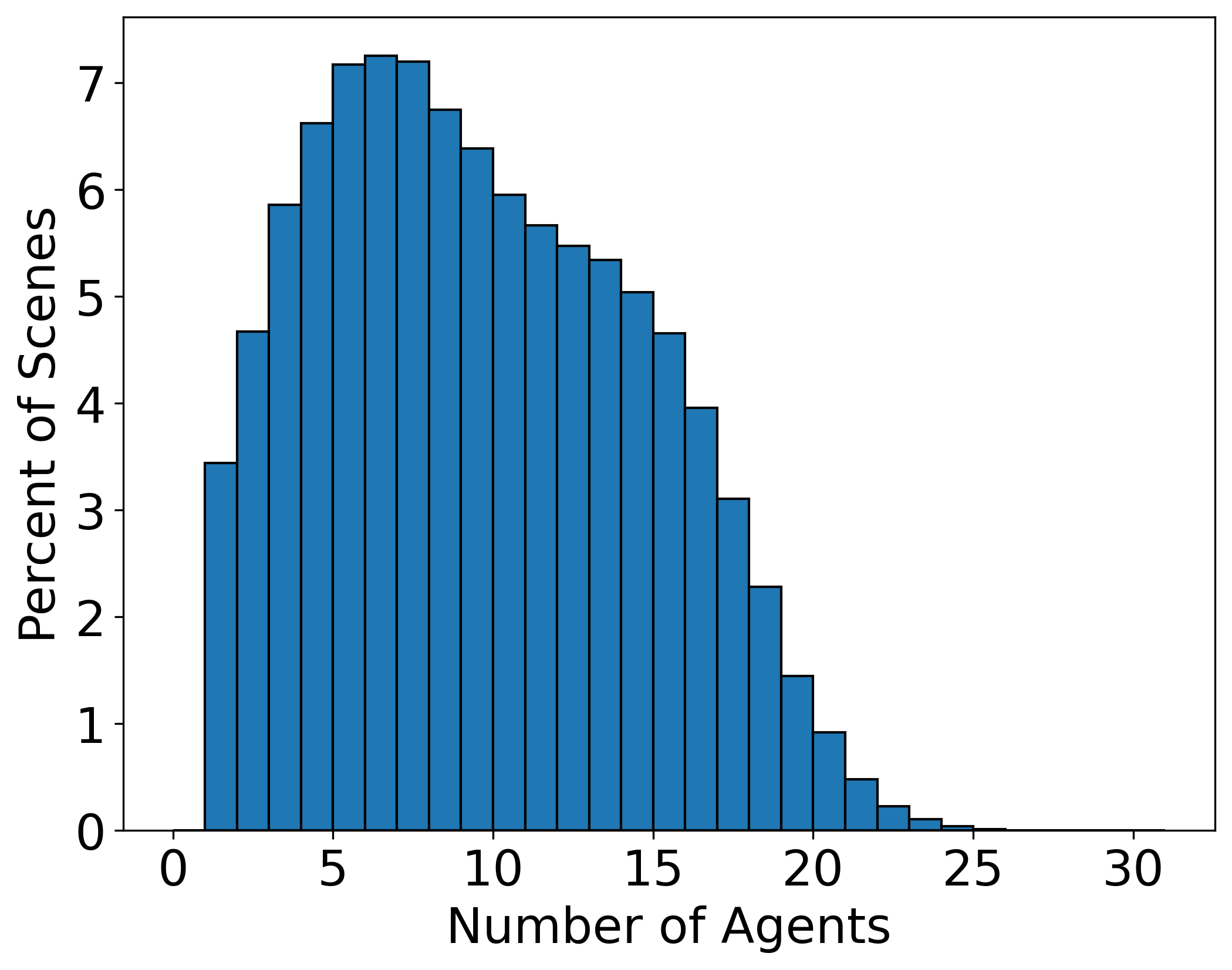}}
  \end{subcaptionbox}
  \hfill
  \begin{subcaptionbox}{Speed distribution of vehicles\label{fig:sub2}}[0.32\textwidth]
    {\includegraphics[width=\linewidth]{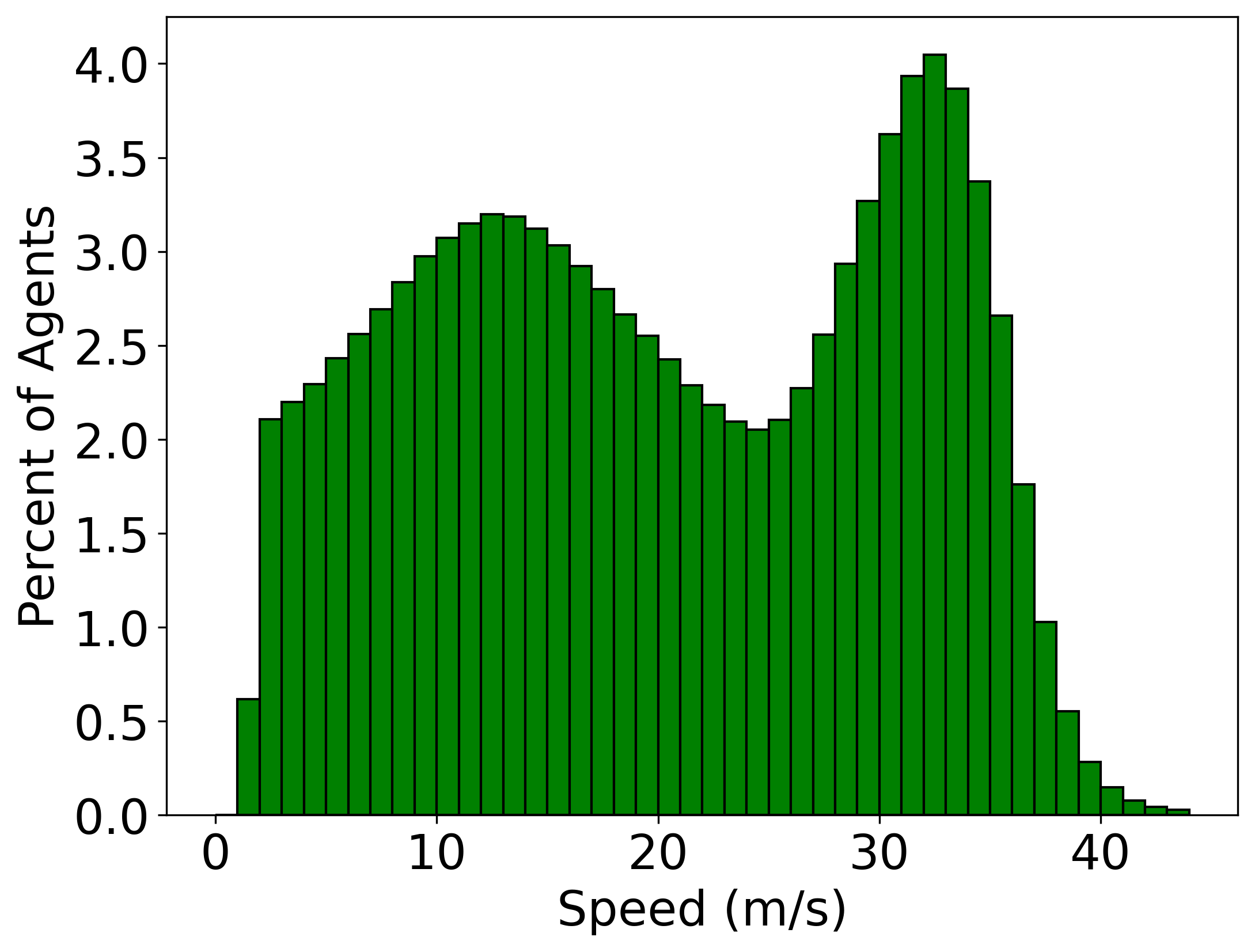}}
  \end{subcaptionbox}
  \hfill
  \begin{subcaptionbox}{Ground truth trajectory of each vehicle starting at origin, color coded based on speed\label{fig:sub3}}[0.33\textwidth]
    {\includegraphics[width=\linewidth]{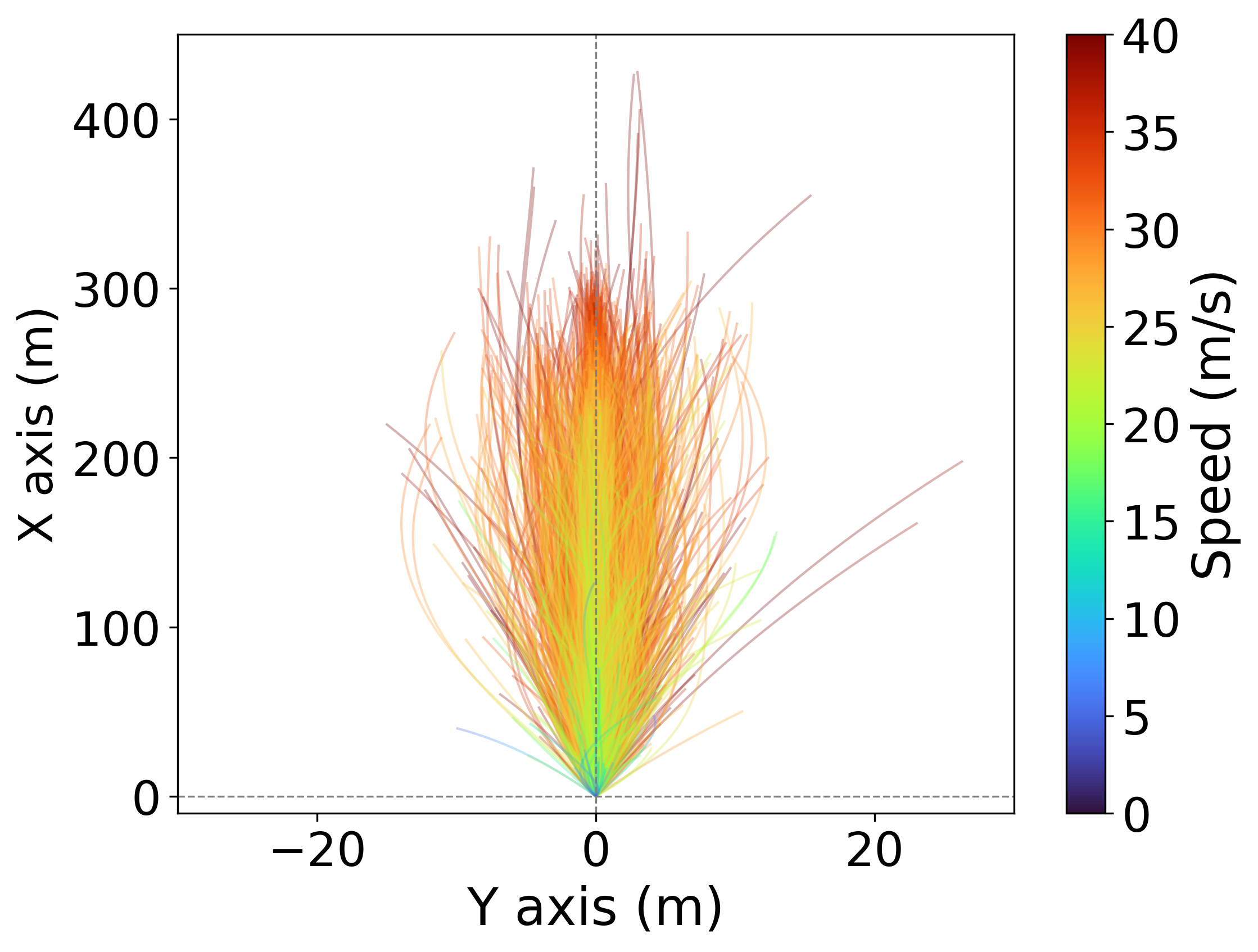}}
  \end{subcaptionbox}
  \caption{Summary statistic visualizations of the I24-MSD Dataset scenarios}
  \label{fig:summary-data}
\end{figure*}

As a summary of the dataset, we borrow the AV traffic simulation dataset comparison from Ettinger et al.~\cite{ettinger2021large} and extend it with I24-MSD dataset statistics in Table~\ref{dataset-comparison}.  Additionally, Figure~\ref{fig:summary-data} presents visualizations of key scenario characteristics in I24-MSD, including agent count distribution, speed distribution, and vehicle trajectory distribution.

\section{Noise-aware Optimizations of Generative Traffic Models}

Next, we look at optimizing generative models for microscopic traffic simulation with noise-aware loss functions. To this end, 
we adapt the state-of-the-art SMART model~\cite{wu2024smart}, which is widely adopted in AV traffic simulation, to better handle imperfections present in microscopic traffic simulation data. SMART utilizes a GPT-style, decoder-only Transformer to model vehicle motion as a next-token prediction task, where each token encodes a relative change in the vehicle’s state—specifically, the relative \(x\), relative \(y\), and relative heading from the current time step. The objective is to autoregressively predict the next motion token for each vehicle, conditioned on its one-second-long past trajectory.

We take inspiration from the computer vision and large language model training~\cite{cambrin2024beyond}, where handling noise and imperfections in data is a common challenge. We train and evaluate SMART on the I24-MSD dataset using one standard loss function (non-noise-specific) and three noise-aware loss functions- all of which have been proposed or used in prior work to improve generalization under data imperfections- that are often used in these communities, specifically in next-token-prediction tasks.

These noise-aware loss functions aim to address a few but not all central data-related challenges. The first challenge is behavioral imbalance. We observe that a majority of vehicles in the dataset tend to travel in relatively straight paths, while maneuvers such as lane changes or deceleration in stop-and-go traffic are comparatively rare. Yet, these rarer behaviors are crucial for simulation fidelity. This results in a class imbalance where dominant behaviors—such as free-flow driving— imbalance the dataset.

which can cause generative models to underperform on less frequent but critical behaviors. The second challenge is label noise and jitter. As detailed in Section~\ref{noise-sources}, sensor noise, tracking inconsistencies, and projection errors can introduce jitter and label noise (in the context of SMART, a token index is the label) into the ground truth trajectories. Naively treating this data as noise-free can degrade learning outcomes and reduce the robustness of the learned model.

To address these issues, we benchmark SMART trained with the following loss functions:

\subsubsection{Cross-Entropy Loss}

Cross-entropy loss is the standard loss function used in most of the token-prediction-based generative traffic simulation methods~\cite{wu2024smart, philion2023trajeglish}. It quantifies the dissimilarity between the predicted probability distribution \(\hat{\mathbf{p}} \in \mathbb{R}^C\) over \(C\) tokens and the one-hot encoded ground truth tokens \(\mathbf{y} \in \{0, 1\}^C\). It is defined as:

\begin{equation}
\mathcal{L}_{\text{CE}}(\mathbf{y}, \hat{\mathbf{p}}) = - \sum_{i=1}^C y_i \log(\hat{p}_i)
\label{ce-loss}
\end{equation}

While effective for clean and balanced data, cross-entropy is sensitive to both token noise and class imbalance.

\subsubsection{Cross-Entropy with Label Smoothing}

Label smoothing is a regularization technique that replaces the hard one-hot token vector with a soft target distribution~\cite{muller2019does}. For a smoothing parameter \(\varepsilon \in [0, 1]\), the target becomes:

\[
y_i^{\text{smooth}} = (1 - \varepsilon) \cdot y_i + \frac{\varepsilon}{C}
\]

The smoothed cross-entropy loss is then:

\begin{equation}
\mathcal{L}_{\text{LS}}(\mathbf{y}, \hat{\mathbf{p}}) = - \sum_{i=1}^C \left[ (1 - \varepsilon) y_i + \frac{\varepsilon}{C} \right] \log(\hat{p}_i)
\label{label-smoothing}
\end{equation}

This approach discourages overconfident predictions and provides moderate robustness to noisy tokens by softening incorrect targets.

\subsubsection{Focal Loss}

Focal loss~\cite{lin2017focal} is designed to address class imbalance by down-weighting well-classified examples and focusing learning on harder, misclassified ones. For a tunable focusing parameter \(\gamma > 0\), focal loss is given by:

\begin{equation}
\mathcal{L}_{\text{Focal}}(\mathbf{y}, \hat{\mathbf{p}}) = - \sum_{i=1}^C y_i \, (1 - \hat{p}_i)^\gamma \log(\hat{p}_i)
\label{focal-loss}
\end{equation}

\noindent When \(\gamma = 0\), this reduces to standard cross-entropy. Higher \(\gamma\) increases the focus on misclassified samples, making it particularly useful in imbalanced settings.

\begin{table*}[h]
    \centering
    \begin{tabular}{|c|c|c|c|c|c|}
        \hline
        Method & Realism~(↑) & Kinematic~(↑) & Interactive~(↑) & Map-Based~(↑) & minADE~(↓) \\
        \hline \hline
        IDM & 0.7001 & \textbf{0.7592} & 0.8192 & 0.5365 & 4.0632 \\
        \hline
        Constant Speed & 0.6891 & 0.7581 & 0.7904 & 0.5429 & 4.2243 \\
        \hline
        SMART (CE) & 0.7698 & 0.7353 & 0.8253 & 0.7183 & 2.0083\\
        \hline
        SMART (CE + LS) & \textbf{0.7922} & 0.7406 & \textbf{0.8300} & \textbf{0.7731} & \textbf{1.3352}\\
        \hline
        SMART (Focal) & 0.7896 & 0.7386 & \textbf{0.8300} & 0.7667 & 1.4417\\
        \hline
        SMART (SCE) & 0.7837 & 0.7382 & 0.8281 & 0.7526 & 1.5929\\
        \hline
    \end{tabular}
    \caption{
    Performance comparison of noise-aware optimization techniques on the I-24 MOTION Scenario dataset.
    \newline
    \textit{CE}: Cross-entropy, \textit{CE + LS}: Cross-entropy with label smoothing, \textit{Focal}: Focal loss, \textit{SCE}: Symmetric cross-entropy.
    }
    \label{tab:results}
\end{table*}

\subsubsection{Symmetric Cross-Entropy Loss}

Symmetric cross-entropy (SCE)~\cite{wang2019symmetric} combines standard cross-entropy with reversed cross-entropy to enhance robustness to token noise. The reversed cross-entropy term penalizes overly confident incorrect predictions. The SCE loss is defined as:

\begin{equation}
\mathcal{L}_{\text{SCE}}(\mathbf{y}, \hat{\mathbf{p}}) = \alpha \, \mathcal{L}_{\text{CE}}(\mathbf{y}, \hat{\mathbf{p}}) + \beta \sum_{i=1}^C \hat{p}_i \log(y_i + \eta)
\label{sce-loss}
\end{equation}

\noindent where \(\alpha\) and \(\beta\) are weighting hyperparameters, and \(\eta\) is a small constant to ensure numerical stability. This dual-term formulation allows SCE to maintain good performance under both clean and noisy conditions.

\textbf{Remark}: The objective of this section is not to introduce new noise-aware loss functions, but to explore and repurpose existing ones from other domains such as computer vision. Our goal is to evaluate their effectiveness in the context of microscopic traffic simulation, thereby demonstrating the importance of handling data noise. We also hope that this analysis serves as a strong baseline and motivation for future research in this area.

\section{Experiments and Results}

\subsection{Experiment setup}

In our experiments, we utilize the default I24-MSD dataset, which supports up to 32 agents per scenario, which aligns with AV traffic simulation. Each scenario includes one second of driving history recorded at 10 Hz (i.e., 0.1-second intervals), and we set the prediction horizon to 8 seconds, corresponding to 80 future steps. For the SMART model, we use a token vocabulary size of 512, derived using the k-disks algorithm~\cite{philion2023trajeglish} and 8 million learnable parameters.

To compare the performance of noise-aware optimizations, we compare against two widely used baseline algorithms in microscopic traffic simulation and the SMART model with standard cross-entropy loss.

\begin{itemize}
\item \textbf{IDM (Intelligent Driver Model)}\cite{treiber2013traffic}: A classic car-following model commonly used in microscopic traffic simulation. We calibrate the IDM parameters specifically for the Interstate 24 driving conditions.
\item \textbf{Constant Speed}: This baseline assumes each vehicle maintains the velocity observed at the final timestep of the one-second history. Given that I24-MSD captures freeway driving, this model can be specifically effective in free-flow traffic scenarios.
\item \textbf{SMART}\cite{wu2024smart}: We also evaluate the SMART with cross-entropy loss that is not specifically optimized for noise.
\end{itemize}

In evaluations, we adopt the same metrics used in AV traffic simulation, specifically following the evaluation protocol of the Waymo Sim Agents Challenge~\cite{montali2023waymo}. In this framework, agents are expected to stochastically generate realistic driving scenarios. A realistic simulation is defined as one that reflects the true distribution of driving scenarios observed in the real world. While the exact analytic form of this distribution is unknown, we have access to empirical samples from it through the I24-MSD dataset. Then, following the Waymo Sim Agent Challenge, we calculate the approximate negative log-likelihood of real-world samples under the distribution induced by the simulated samples. For further details, we refer the reader to the Waymo Sim Agents Challenge~\cite{montali2023waymo}. In summary, evaluation is conducted across four key dimensions: realism, kinematics, interactivity, and map-based compliance.

In the loss functions, we use $\epsilon=0.1$ for label smoothing, $\gamma = 2$ in focal loss, and $\alpha = 1$, $\beta = 0.13$, and $\eta=0.0004$ in symmetric cross entropy loss. 

\subsection{Results}

Table~\ref{tab:results} presents the evaluation results for the baselines, the standard SMART model trained with cross-entropy loss, and several SMART variants trained with cross-entropy with label smoothing, focal loss, and symmetric cross-entropy loss. The kinematic, interactive, map-based, and minADE metrics are constructed from component metrics following Montali et al.\cite{montali2023waymo}, while the realism metric is a meta-metric. Due to space constraints, we refer the reader to Montali et al.\cite{montali2023waymo} for a detailed description of these component metrics.

As shown in Table~\ref{tab:results}, all SMART variants outperform the IDM and Constant Speed baselines, highlighting the expressive power of generative models in microscopic traffic simulation. Among the variants, SMART trained with cross-entropy and label smoothing achieves the best overall performance, suggesting that label smoothing helps mitigate overfitting to noisy or ambiguous tokens. Additionally, all noise-aware training loss functions—label smoothing, focal loss, and symmetric cross-entropy—outperform standard cross-entropy, underscoring the benefit of accounting for token noise during training.

\section{Conclusion and Future Work}

In this work, we introduce the I-24 MOTION Scenario Dataset (I24-MSD), a scenario-based vehicle trajectory dataset collected using infrastructure-based cameras, aimed at advancing generative microscopic traffic simulation. Through empirical studies, we show that explicitly accounting for noise and imperfections in training data leads to more accurate and realistic simulations. To account for these imperfections, we explore the use of noise-aware loss functions during model training. With the release of I24-MSD, we hope to inspire further research in generative microscopic traffic simulation with techniques like reinforcement learning-based closed-loop fine-tuning, the development of noise-aware model architectures, and other learning techniques to enhance simulation fidelity. We hope this work establishes a foundation for future progress in generative microscopic traffic simulation, and hence more broadly, in intelligent transportation research and practice.

\section{Acknowdlgement}

The authors would like to thank Cameron Hickert, Zhengbing He, Han Zheng, and Tsung-Han Lin for constructive discussions and their feedback on this work. The authors also thank Derek Gloudemans and Gergely Zach\'ar for providing the data to create the vectorized road maps of Interstate 24 and the coordination system transformation code.

\bibliographystyle{unsrt}
\bibliography{references}

\end{document}